\documentclass{svproc}
\usepackage{url}
\usepackage{graphicx}
\usepackage{braket}
\usepackage{amsmath}
\usepackage{amssymb}
\usepackage{xcolor}

\newcommand{\WignerSIXj}[6]
	{
	\left\{
		\begin{array}{ccc}
			#1 & #2 & #3 \\
   			#4 & #5 & #6
		\end{array}
	\right\}
	}

\newcommand{\WignerNINEj}[9]
	{
	\left\{
		\begin{array}{ccc}
			#1 & #2 & #3 \\
   			#4 & #5 & #6 \\
   			#7 & #8 & #9
		\end{array}
	\right\}
	}

\newcommand{\half}[0]
	{
	  \frac{ 1 }{ 2 }
	}

\begin{document}
\mainmatter              
\title{New symmetry-adapted \textit{ab initio} approach to nuclear reactions for intermediate-mass nuclei}
\titlerunning{New symmetry-adapted \textit{ab initio} approach to nuclear reactions}  
%
\author{Alexis Mercenne\inst{1} \and Kristina D. Launey\inst{1} \and Jutta E. Escher\inst{2} \and Tomas Dytrych\inst{1,3} \and Jerry P. Draayer\inst{1}}
\authorrunning{Alexis Mercenne et al.} 
%
%
\institute{Department of Physics and Astronomy, Louisiana State University, Baton Rouge, LA 70803, USA,\\
\email{amercenne1@lsu.edu}
\and
Lawrence Livermore National Laboratory, Livermore, CA, 94550, USA
\and
Nuclear Physics Institute, Academy of Sciences of the Czech Republic, 25068 {\v{R}}e{\v{z}}, Czech Republic}

\maketitle              

\begin{abstract}
  {
  With a view toward describing reactions of intermediate-mass nuclei from first principles, we present first results for the norm and Hamiltonian overlaps (kernels) for the p-${ \alpha }$, p-${ {  }^{ 16 } }$O and p-${ {  }^{ 20 } }$Ne cluster systems using realistic nucleon-nucleon interactions. 
  This is achieved in the framework of a new \textit{ab initio} approach that combines the symmetry-adapted no-core shell model (SA-NCSM) with the resonating group method (RGM).
 In this model, a physically relevant basis based on the SU(3) symmetry is used. 
 The structure of the clusters is provided by the \textit{ab initio} SA-NCSM, which enables the description of spatially enhanced nuclear configurations and heavier nuclei, by exploiting symmetries known to dominate in nuclei.
 Here, we discuss the applicability  and efficacy of this approach.
  }
\end{abstract}

\noindent \textbf{Introduction. -- }
{
  \textit{Ab initio} descriptions of reactions of nuclei heavier than $^{16}$O remain a challenge in nuclear physics.
  Their theoretical and experimental study is of utmost importance to identify various quantum mechanisms that can explain the complexity of nuclei.
  In addition, many simulations of astrophysical phenomena are very sensitive to nuclear reaction cross sections.
  For example, simulations of X-ray burst nucleosynthesis have been found to be sensitive to several nuclear reaction rates for intermediate- and medium-mass nuclei \cite{cyburt_2016}, pointing to the need for accurate cross sections.
  For theoretical predictions, this level of accuracy can be achieved through an \textit{ab initio} description of nuclear reactions.
  Recent progresses in \textit{ab initio} nuclear theory using QCD-inspired realistic interactions along with the continuous improvement of high performance computing have given the necessary tools to theoretical approaches such as the no-core shell model (NCSM) to provide an \textit{ab initio} description of the structure of light nuclei \cite{navratil_2000,barrett_2013}.
  Its recent implementation within the RGM \cite{RGM_tang_1978}, the NCSM/RGM, has allowed a microscopic study of nuclear reactions \cite{quaglioni_2008,quaglioni_2009,baroni_2013}, pursuing the long-standing goal of unifying nuclear structure and reactions.
  Recently, it has been demonstrated that the SA-NCSM \cite{dytrych_2013,launey_2016_review}, which employs a physically relevant basis, can use drastically reduced configuration spaces with practically the same accuracy of results, and has been successfully applied up to medium-mass nuclei \cite{LauneySOTANCP42018,DraayerDL16}.
  Motivated by the need for calculated nuclear cross sections in experimental research and astrophysical studies, and following the success of the NCSM/RGM for light nuclei, we combine the SA-NCSM with the RGM. 
  As a first step, we focus on reactions of two clusters, in which the projectile is a nucleon. \\
}

\noindent \textbf{\textit{Ab initio} symmetry-adapted framework for nuclear reactions. --}
{
In the RGM framework, the nucleons are organized within different groups, or clusters, ``resonating'' through the intercluster exchange of nucleons. 
This antisymmetrization between the different clusters guarantees the Pauli exclusion principle, which, along with the consideration of the cluster internal structure, is one of the most important features of the approach.
In the case of two clusters, the wave function is written as (in notations of Ref. \cite{quaglioni_2009}): 
  \begin{equation}
    \ket{ { \Psi }^{ { J }^{ \pi } T } } = \sum_{\nu} \int_{r} dr { r }^{ 2 } \frac{ { g }_{ \nu }^{ { J }^{ \pi }T }(r) }{ r } \hat{ \mathcal{A} } \ket{ { \Phi }_{ \nu r }^{ { J }^{ \pi }T } } \;,
    \label{RGM_ansatz}
  \end{equation}
  where the index ${ \nu }$ represents all quantum numbers that define channels and partitions: ${ \nu = \{ (A-a) { \alpha }_{ 1 } { I }_{ 1 } { T }_{ 1 } ; a { \alpha }_{ 2 } { I }_{ 2 } { T }_{ 2 } ; \ell s \} }$, and the cluster states are defined as ${ \ket{ { \Phi }_{ \nu r }^{ { J }^{ \pi }T } } = { \left[ { \left( \ket{ (A-a) { \alpha }_{ 1 } { I }_{ 1 } { T }_{ 1 } } \otimes \ket{ a { \alpha }_{ 2 } { I }_{ 2 } { T }_{ 2 } } \right) }^{ (sT) } \!\! \times { Y }_{ \ell } ({ \hat{ r } }_{ A-a,a }) \right] }^{ ({ J }^{ \pi }T) } \frac{ \delta(r - { r }_{ A-a,a }) }{ r { r }_{ A-a,a } } }$. 
  The amplitudes ${ { g }_{ \nu }^{ { J }^{ \pi }T }(r) }$ 
describe the relative motion between the target and the projectile for all channels ${ \nu }$, and the cross section can be extracted from their asymptotic behavior.
The ${ { g }_{ \nu }^{ { J }^{ \pi }T } (r) }$  functions are the solutions to the Schr\"odinger equation: 
  \begin{equation}
    \sum_{\nu} \int dr { r }^{ 2 } \left[ { H }_{ \nu' \nu }^{ { J }^{ \pi }T } (r,r') - E { N }_{ \nu'\nu }^{ { J }^{ \pi }T }(r',r) \right] \frac{ { g }_{ \nu }^{ { J }^{ \pi }T } (r) }{ r } = 0 \; .
    \label{RGM_equations}
  \end{equation}
  Here, the Hamiltonian ${ { H }_{ \nu'\nu }^{ { J }^{ \pi }T } (r',r) }$ and norm ${ { N }_{ \nu' \nu }^{ { J }^{ \pi }T }(r',r) }$ kernels are expressed as ${ \bra{ { \Phi }_{ \nu' r' }^{ { J }^{ \pi }T } } \hat{ \mathcal{A} } \hat{ O } \hat{ \mathcal{A} } \ket{ { \Phi }_{ \nu r }^{ { J }^{ \pi }T } } }$ with $\hat{ O }$ being the identity and the Hamiltonian operator, respectively, and where ${ \hat{ \mathcal{A} } }$ is the antisymmetrizer ensuring the Pauli exclusion principle. 
  The kernels are computed using the wave functions of the clusters.
  Eq. (\ref{RGM_equations}) can then be solved using an ${ R }$-matrix approach \cite{descouvemont_Rmatrix,descouvemont_code}.

  An \textit{ab initio} application of this approach is the NCSM/RGM \cite{quaglioni_2009}, which uses NCSM wave functions and realistic interactions.
  However, the method becomes numerically challenging for heavier systems due to the size and complexity of the configuration space. 
  We address the limitation of the NCSM/RGM by combining the SA-NCSM with the RGM formalism, where the former allows for the calculation of the intermediate mass wavefunctions required by the RGM.

  In the SA-NCSM, the microscopic many-body basis is based on the spherical harmonic oscillator single-particle basis, and labeled by irreducible representations according to the group chain: 
  \begin{equation}
     {\text{SU}(3)}_{ (\lambda \mu) } \underset{\kappa}{\supset} {\text{SO}(3)}_{ L } \supset {\text{SO}(2)}_{ { M }_{ L } }.
    \label{}
  \end{equation}
  Consequently, for any given total spin ${ J }$ and its projection ${ M }$, the wave function of a nucleus will be described within a basis ${ \{ \ket{ { \alpha }_{ i } ({ \lambda }_{ i } { \mu }_{ i }) { \kappa }_{ i } ({ L }_{ i }{ S }_{ i }) J M } \} }$ with each component weighted by a coefficient ${ { C }_{ i } }$. 
  Here ${ { \alpha }_{ i } }$ represents additional quantum numbers needed to enumerate the complete shell-model space. 
  
  In the SA-RGM, the channels are defined by coupling each component of the SA-NCSM wave functions between the projectile and the target.
  Consequently, the channels with good SU(3) spin and isospin quantum numbers are given in the case of one nucleon projectile as:

  \begin{equation}
    \ket{ { \Phi }_{ \gamma n }^{ \rho(\lambda \mu)\kappa(LS)JM T { M }_{ T } } } = { \{ \ket{ { \alpha }_{ 1 } ({ \lambda }_{ 1 } { \mu }_{ 1 }) { S }_{ 1 } { T }_{ 1 } } \otimes \ket{ (n0) \frac{ 1 }{ 2 } \frac{ 1 }{ 2 } } \} }^{ \rho (\lambda \mu) \kappa(LS) JM T { M }_{ T } } \; ,
    \label{}
  \end{equation}
  where the index ${ \gamma \equiv \{ { \alpha }_{ 1 } ({ \lambda }_{ 1 } { \mu }_{ 1 }) { S }_{ 1 } { T }_{ 1 } ; (n 0) \frac{ 1 }{ 2 } \frac{ 1 }{ 2 } \} }$ labels the channel basis, with ${ (n0) }$ representing the SU(3) labels of the projectile with spin  $\frac{ 1 }{ 2 }$ and isospin $\frac{ 1 }{ 2 }$.
  Note that there is no dependence on the orbital momentum of the target and the projectile. 

  In this basis, the exchange matrix, which ensures the antisymmetrization in the kernels, has the following form (in conventional notations \cite{draayer_1973}): 
  \begin{align}
    & \bra{ { \Phi }_{ \gamma'n' }^{ \rho'(\lambda' \mu')\kappa'(L'S')JMT{ M }_{ T } } } { \hat{ P } }_{ A,A-1 } \ket{ { \Phi }_{ \gamma n }^{ \rho(\lambda \mu) \kappa(LS) JM T { M }_{ T } } } \nonumber \\
    & = \frac{ 1 }{ A - 1 } { \delta }_{ \rho \rho' } { \delta }_{ (\lambda \mu)(\lambda' \mu') } { \delta }_{ \kappa \kappa' } { \delta }_{ LL' } { \delta }_{ SS' } \sum_{ \substack{ \tau { \rho }_{ o } ({ \lambda }_{ o } { \mu }_{ o }) \\ { S }_{ o } \bar{\rho} } } { \Pi }_{ \tau { S }_{ o } { S }_{ 1 }' { T }_{ 1 }' } { (-1) }^{ n + n' - ({ \lambda }_{ o } + { \mu }_{ o }) } \nonumber \\ 
    & \times { (-1) }^{ { T }_{ 1 } + \frac{ 1 }{ 2 } + T' } { (-1) }^{ { S }_{ 1 } +  \frac{ 1 }{ 2 } + S' } \WignerSIXj{ { S }_{ 1 } }{ { S }_{ o } }{ { S }_{ 1 }' }{ \frac{ 1 }{ 2 } }{ S }{ \frac{ 1 }{ 2 } } \WignerSIXj{ { T }_{ 1 } }{ \tau }{ { T }_{ 1 }' }{ \frac{ 1 }{ 2 } }{ T }{ \frac{ 1 }{ 2 } } \nonumber \\ 
    & \times \sqrt{ \frac{ \text{dim}({ \lambda }_{ o } { \mu }_{ o }) }{ \text{dim}(n 0) } } \text{U}\left[ ({ \lambda }_{ 1 } { \mu }_{ 1 }) ({ \lambda }_{ o } { \mu }_{ o }) (\lambda' \mu') (n'0) ; ({ \lambda }_{ 1 }' { \mu }_{ 1 }') \bar{\rho} \rho' (n0) { \rho }_{ o } \rho'' \right] \nonumber \\
    & \times \bra{ { \alpha }_{ 1 }' ({ \lambda }_{ 1 }' { \mu }_{ 1 }') { S }_{ 1 }' { T }_{ 1 }' } || { \{ { a }_{ (n 0) \frac{ 1 }{ 2 } \frac{ 1 }{ 2 } }^{ \dagger } \otimes { \tilde{ a } }_{ ( \tilde{ 0 n' } ) \frac{ 1 }{ 2 } \frac{ 1 }{ 2 } } \} }^{ { \rho }_{ o } ({ \lambda }_{ o } { \mu }_{ o }) { S }_{ o } \tau } || { \ket{ { \alpha }_{ 1 } ({ \lambda }_{ 1 } { \mu }_{ 1 }) { S }_{ 1 } { T }_{ 1 } } }_{\bar{\rho} } .
    \label{FULL_SU3_CHANNELS}
  \end{align}
  Clearly, the presence of the delta Kronecker functions in Eq.(\ref{FULL_SU3_CHANNELS}) makes the exchange matrix diagonal within this SU(3) basis, allowing for several numerical simplifications \cite{hecht_1977}. 
  Furthermore, matrix calculations avoid complications of dealing with the orbital momentum, which is introduced at the very last step of the calculation, 
  for input to the $R$-matrix approach. Namely, we can retrieve the partial-wave expansion
  \begin{align}
    \ket{ { \Phi }_{ \nu n }^{ JMT{ M }_{ T } } } & = \sum_{i} { C }_{ i } \sum_{ \substack{ j \rho(\lambda \mu) \\ \kappa LS } } { \Pi }_{ { I }_{ 1 } L S j } \bra{ ({ \lambda }_{ 1 }^{ i } { \mu }_{ 1 }^{ i }) { \kappa }_{ 1 }^{ i } { L }_{ 1 }^{ i } ; (n0) 0 \ell } { \ket{ (\lambda \mu)K L } }_{ \rho } \nonumber \\ 
    & \times { (-1) }^{ { I }_{ 1 } + J + j } { \Pi }_{ sj } \WignerSIXj{ { I }_{ 1 } }{ \half }{ s }{ \ell }{ J }{ j } \WignerNINEj{ { L }_{ 1 }^{i} }{ { S }_{ 1 }^{ i } }{ { I }_{ 1 } }{ \ell }{ \frac{ 1 }{ 2 } }{ j }{ L }{ S }{ J } \ket{ { \Phi }_{ { \gamma }_{ i } n }^{ \rho(\lambda \mu)\kappa(LS)JM T { M }_{ T } } } 
    \label{fromSU3toSU2}
  \end{align}
  and calculate the norm ${ { N }_{ \nu' \nu }^{ { J }^{ \pi } T } (r',r) }$ using the formula of Ref. \cite{quaglioni_2009}.
  Note that the summation over ${ i }$ represents the expansion of the target wave function in terms of the SU(3) basis states, where $i$ is given by ${ \{ { \alpha }_{ 1 }^{ i } ({ \lambda }_{ 1 }^{ i } { \mu }_{ 1 }^{ i }) { \kappa }_{ 1 }^{ i } { L }_{ 1 }^{ i } { S }_{ 1 }^{ i } \} }$.
  The Hamiltonian kernel is calculated straightforwardly using the same procedure, but the details are more complicated and are omitted for brevity here.

  \begin{figure}[t]
    \centering
    \begin{minipage}{0.4\textwidth}
        \centering
	\includegraphics[width=5.5cm]{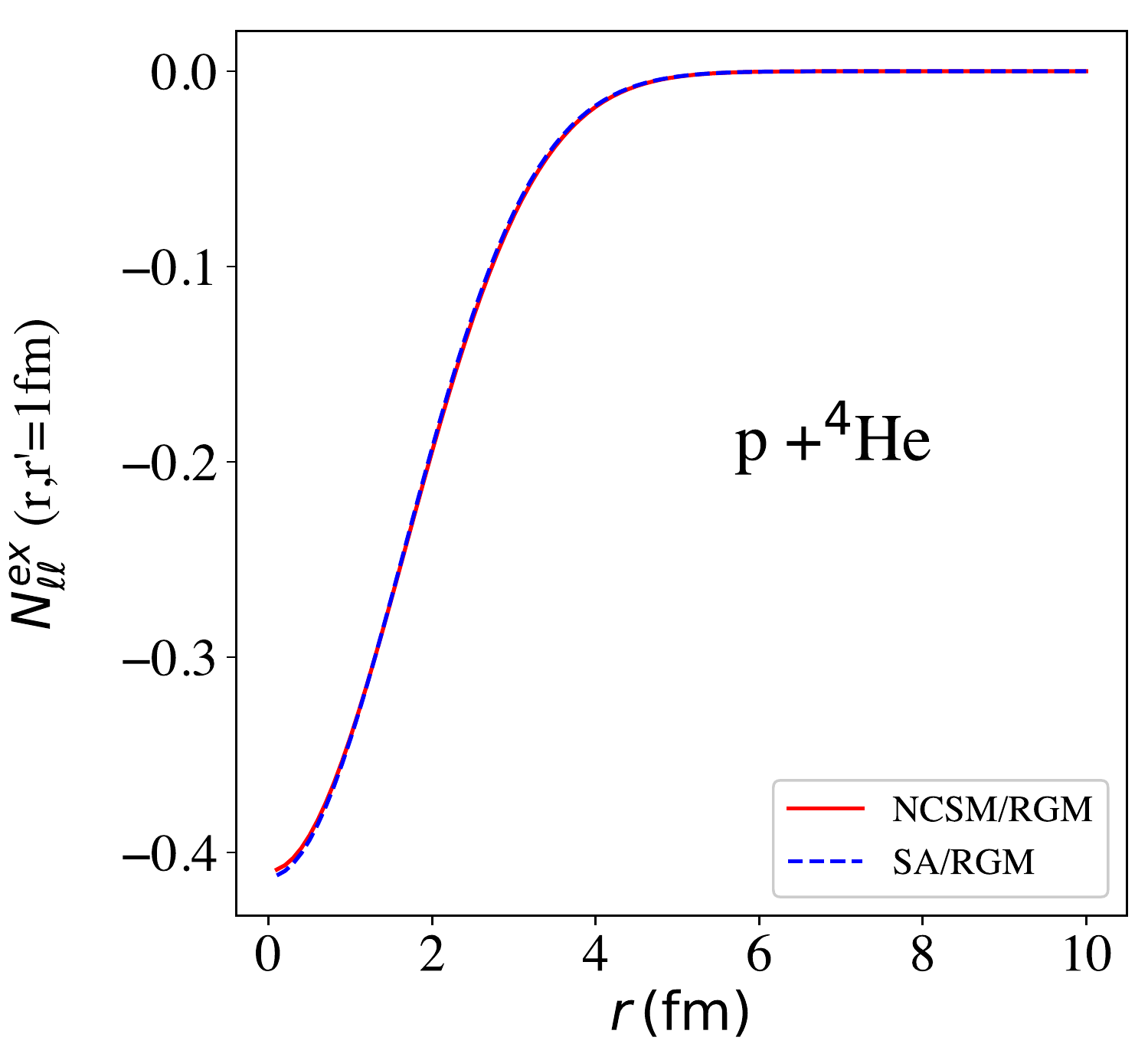} 
    \end{minipage}\hfill
    \begin{minipage}{0.4\textwidth}
        \centering
	\caption{Exchange part of the norm kernel (${ \ell = 0 }$) for p-${ {  }^{ 4 } }$He. The NCSM/RGM calculation was performed using the formalism of Ref. \cite{quaglioni_2009} and the complete ${ {  }^{ 4 } }$He wave function. The SA-RGM calculation was performed using Eq. (\ref{FULL_SU3_CHANNELS}) and a truncated ${ {  }^{ 4 } }$He wave function, where only SU(3) components with a probability greater than 1\% are selected. Calculations are performed in 4 shells  and for ${ \hbar \Omega = 15 }$ MeV.}
    \end{minipage}
    \label{plot_benchmark}
\end{figure}
}

\vspace{-0.4cm}

  \begin{figure}[h]
    \centering
    \begin{minipage}{0.5\textwidth}
        \centering
	\includegraphics[width=5.5cm]{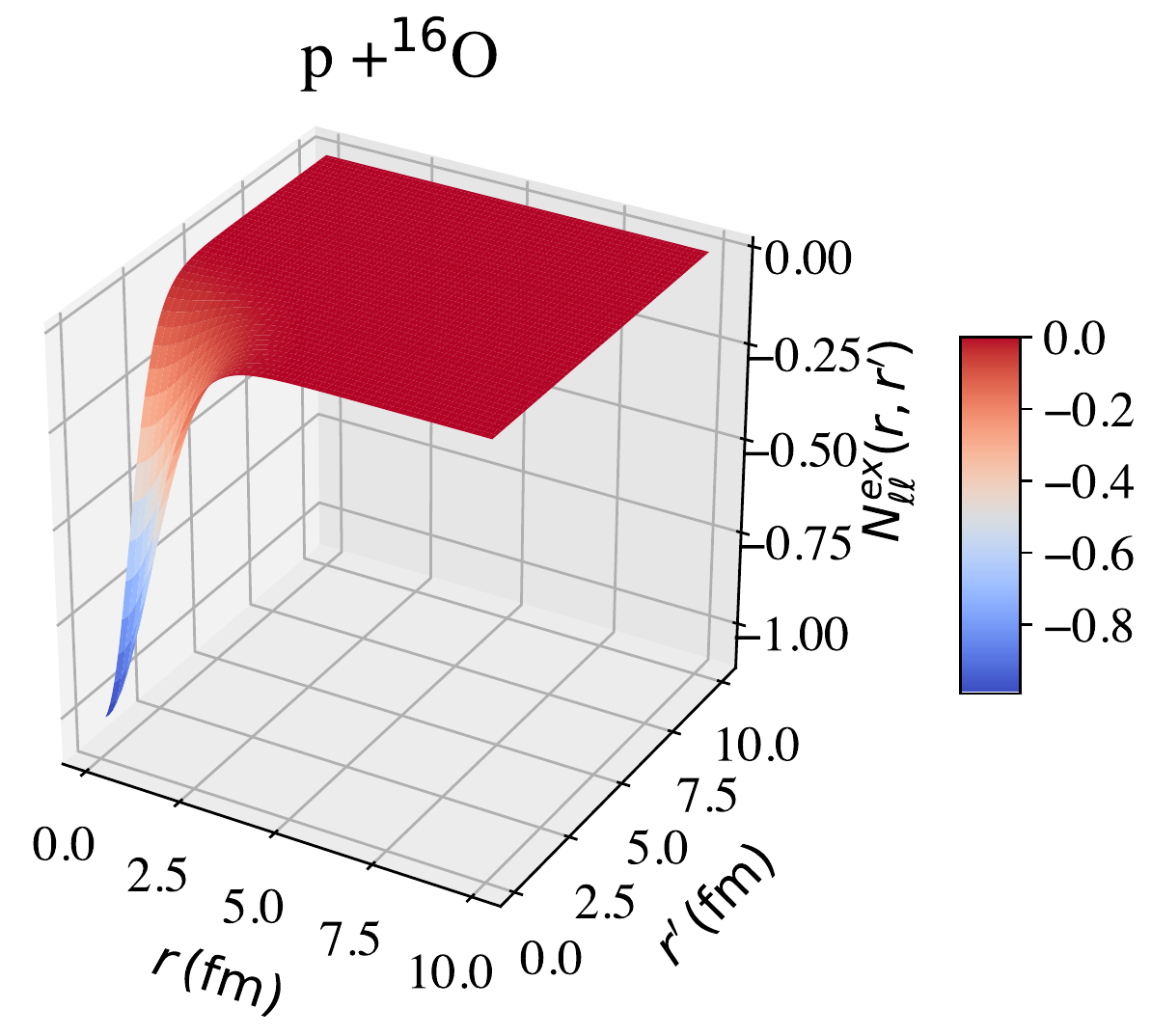} 
    \end{minipage}\hfill
    \begin{minipage}{0.5\textwidth}
        \centering
	\includegraphics[width=5.4cm]{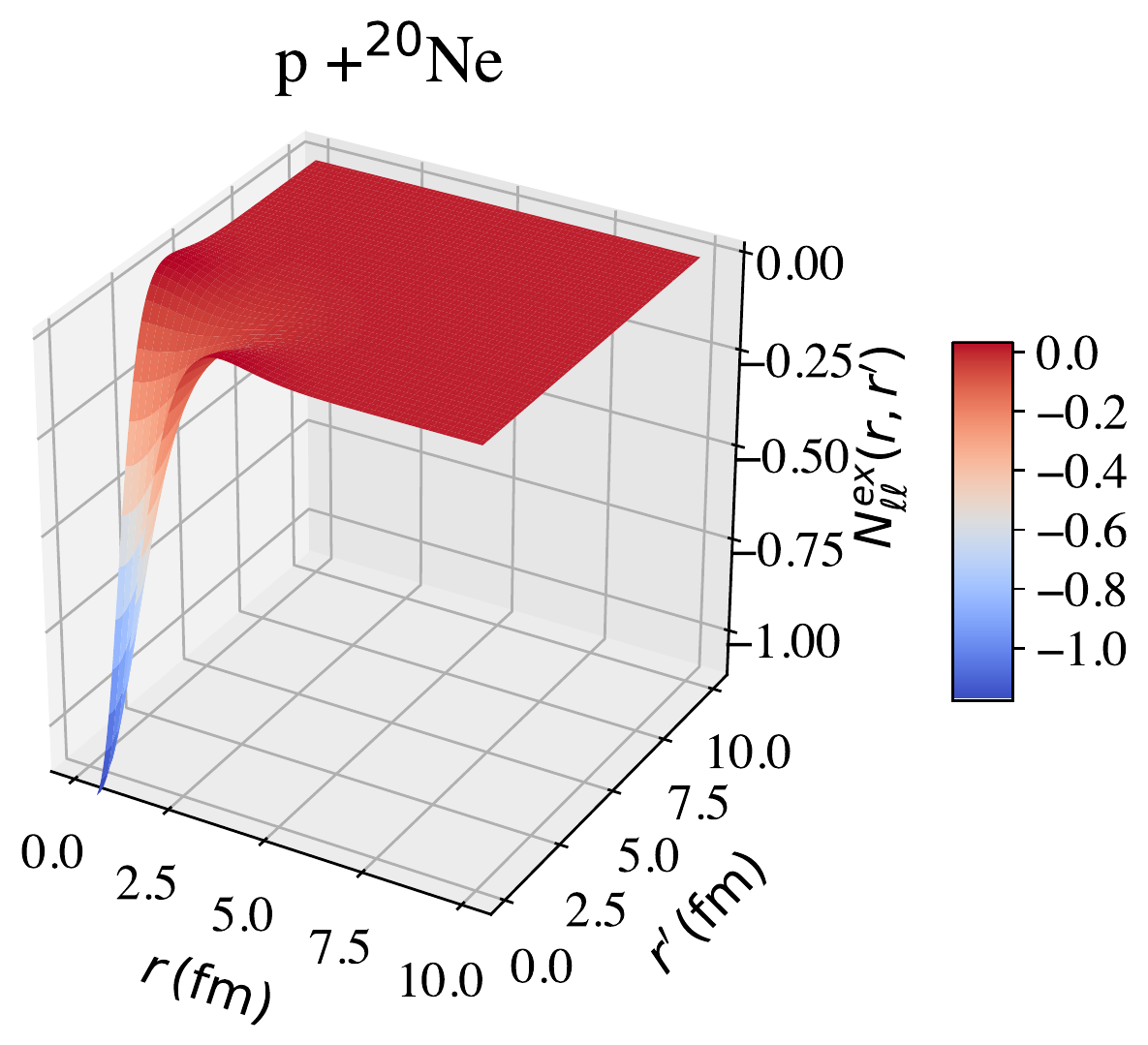} 
    \end{minipage}
    \caption{Exchange part of the norm kernel (${ \ell = 0 }$). The target wave function is calculated using the chiral ${ { \text{NNLO} }_{ \text{sat} } }$ NN in 10 shells (${ \hbar \Omega= 16 }$ MeV) for ${ {  }^{ 16 } }$O, and the chiral ${ { \text{NNLO} }_{ \text{opt} } }$ NN in 13 shells (${ \hbar \Omega = 15 }$ MeV) for ${ {  }^{ 20 } }$Ne, with selected SU(3) configurations that have a contribution greater than 2\%. 
    }
    \label{plot_3D_norm}
\end{figure}

\noindent \textbf{Results. --}
  To demonstrate the efficacy of the approach, we present results  for norm and Hamiltonian kernels for light and intermediate-mass nuclei. 

  SA-NCSM and SA-RGM computations are performed in laboratory coordinates. 
  The center-of-mass (CM) spuriosity is removed for the target wave function. 
  To simplify the calculations the present results are reported for a projectile-target system with the CM included (the removal of the CM is work in progress and is based on an efficient group-theoretical algorithm to be reported in another publication). 
  Nonetheless, this CM effect is expected to be negligible for reactions for one nucleon plus an ${ A \gtrsim 16 }$ target, such as ${ {  }^{ 16 } }$O and ${ {  }^{ 20 } }$Ne.

  \begin{figure}[t]
    \centering
    \begin{minipage}{0.4\textwidth}
        \centering
	\includegraphics[width=5.5cm]{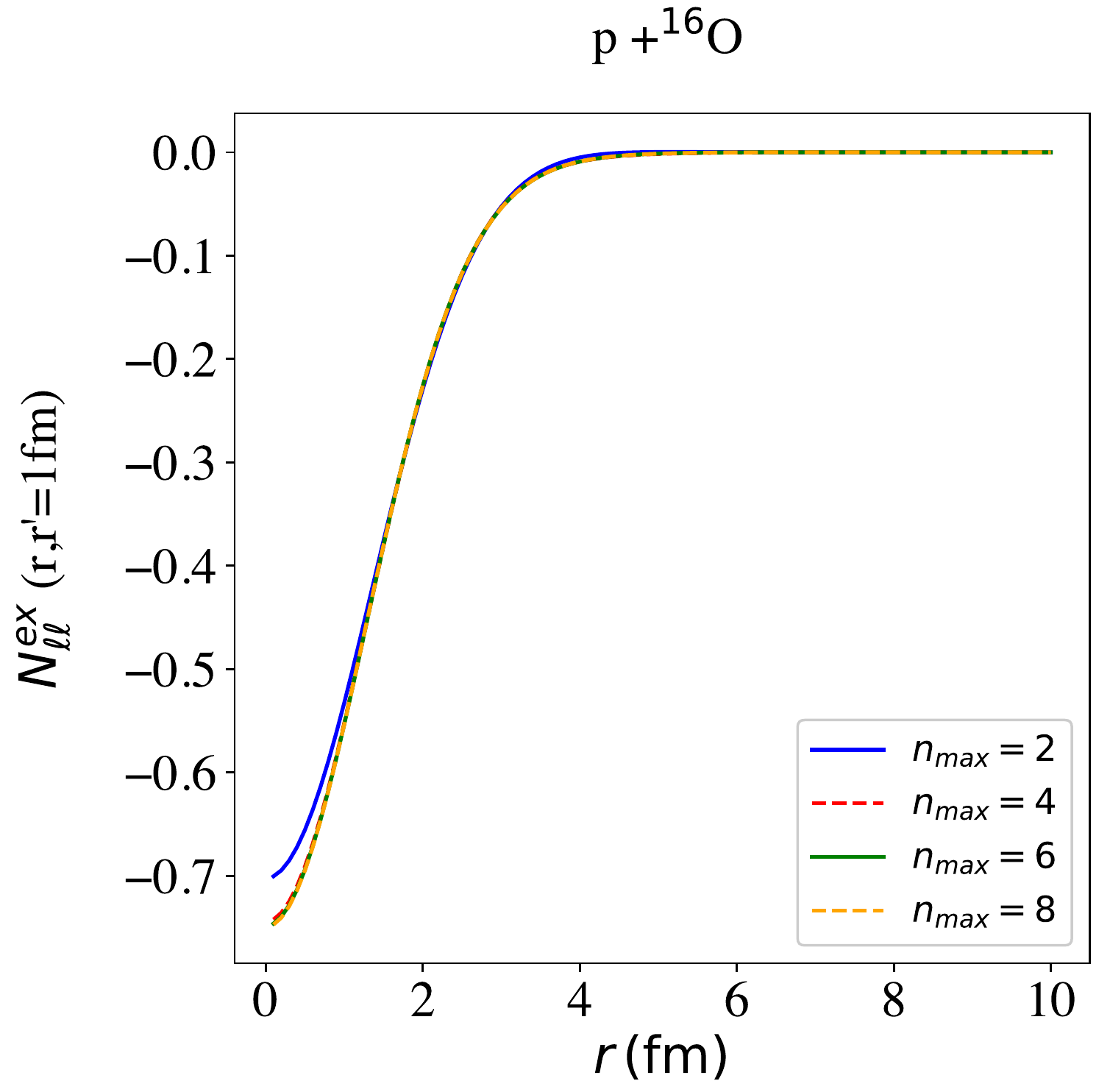} 
    \end{minipage}\hfill
    \begin{minipage}{0.4\textwidth}
        \centering
	\caption{Convergence of the exchange part of the norm (${ \ell = 0 }$) with the allowed number of shells $n_{\rm max}$ for the projectile using SA-RGM. Calculations are described in the caption of Fig. 2. 
	}
    \end{minipage}
    \label{plot_16O}
\end{figure}
  First, we have performed a benchmark calculation for p-${ {  }^{ 4 } }$He, where we compare the exchange part of the norm in laboratory coordinates for the NCSM/RGM approach, according to Eqs. (37) and (50) of Ref. \cite{quaglioni_2009}, and the SA-RGM approach using Eqs. (\ref{FULL_SU3_CHANNELS}) and (\ref{fromSU3toSU2}) (Fig. \ref{plot_benchmark}).
  The SA-RGM result has been obtained using a ${ {  }^{ 4 } }$He wave function truncated to only several SU(3) basis states, and is in excellent agreement with the NCSM/RGM calculation.
  It is important to mention that the SA-RGM approach with the complete  SU(3) wave function provides exactly the same results as in the NCSM/RGM.

  \vspace{-0.5cm}

 \begin{figure}[h!]
    \centering
    \begin{minipage}{0.5\textwidth}
        \centering
	\includegraphics[width=5.5cm]{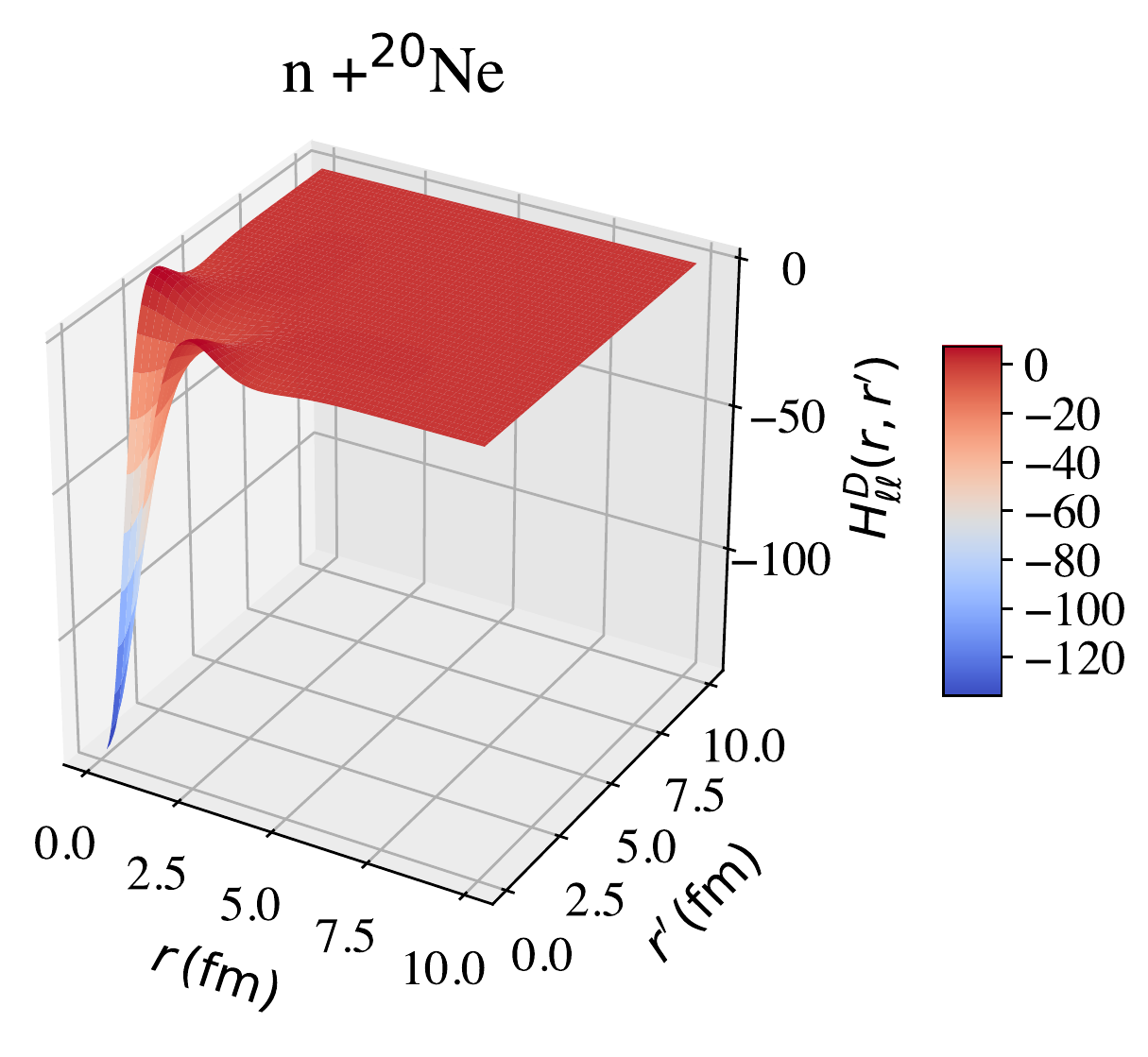} 
    \end{minipage}\hfill
    \begin{minipage}{0.5\textwidth}
        \centering
	\includegraphics[width=5.2cm]{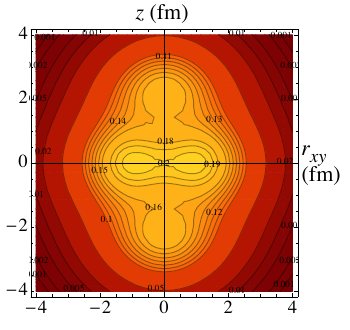} 
    \end{minipage}
    \caption{Left panel: Direct part of the Hamiltonian  kernel (${ \ell = 0 }$) using the same wave function as in Fig. \ref{plot_3D_norm}. Right panel: Corresponding one-body density profile of ${ {  }^{ 20 } }$Ne from the SA-NCSM calculation.}
    \label{plot_direct_H_kernel}
\end{figure}

The first results for the kernels for reactions of intermediate-mass nuclei using various realistic nucleon-nucleon (NN) interactions are now available. 
In these calculations, we use SA-selected model spaces for the target wave functions, as complete SU(3) (equivalent to NCSM) model spaces for a sufficiently large number of shells are prohibitive. 
As an illustrative example, we show the norm kernel for p-${ {  }^{ 16 } }$O with the NNLO$_{\rm sat}$ \cite{n2lo_sat} interaction and p-${ {  }^{ 20 } }$Ne with the NNLO$_{\rm opt}$ \cite{n2lo_opt} interaction (Fig. \ref{plot_3D_norm}). 
As expected, the norm kernel vanishes at large distances, which is consistent with the Pauli principle. 
Results are shown for a model space for the projectile that yields convergence. 
Indeed, we find that the norm kernel converges comparatively quickly for the NNLO$_{\rm sat}$ interaction and, e.g., including up to 4 shells has already yielded a converged norm kernel for  p-${ {  }^{ 16 } }$O (Fig. 3). 

The Hamiltonian kernel provides information on the non-local effective interaction between the projectile and the target for a given channel, and can be studied for intermediate-mass targets in the SA-RGM framework. 
For example, we find that the direct part of the Hamiltonian kernel for the  p$+{ {  }^{20} }$Ne shows a different behavior as compared to doubly-magic systems (Fig. \ref{plot_direct_H_kernel}, left panel).
  The positive peaks occurring around ${ r=3 }$ fm might be related to the intricate structure of ${ {  }^{ 20 } }$Ne that exhibits clustering substructures and enhanced deformation, as shown in the density profile (Fig. \ref{plot_direct_H_kernel}, right panel).
Further investigations of these effective interactions in this region and the role of non-locality are needed, especially in relation to obtaining first-principle optical potentials.

To summarize, the use of a physically relevant basis in the SA-RGM provides a pathway to {\it ab initio} descriptions of nuclear reactions in the intermediate-mass region.
The use of this basis allows several numerical procedures inherent to RGM to be simplified.
The present outcome shows the applicability of the method, including benchmark calculations, convergence properties, and a discussion of non-local inter-cluster effective interactions.

We acknowledge useful discussions with P. Navr\'{a}til and S. Quaglioni. This work was supported by the U.S. National Science Foundation (OIA-1738287, ACI -1713690), the Czech Science Foundation (16-16772S) and under the auspices of the U.S. Department of Energy by Lawrence Livermore National Laboratory under Contract DE-AC52- 07NA27344, with support from LDRD project 19-ERD-017.
In addition, this work benefitted from computing resources provided by LSU ({\tt www.hpc.lsu.edu}), Blue Waters, and NERSC.



\end{document}